%% file: main.tex
\documentclass[runningheads]{llncs}
\usepackage[LNCS,
   key=2021-Heiss-ICSOC-PreProcessing,
   year=2021,
   publication={J. Heiss, A. Busse, and S.Tai, Proceedings of the 19th International Conference on Service-Oriented Computing, ICSCO'21.}
   Preprint,
   nourl,
   nobib
 ]{authorarchive}
\usepackage{graphicx}
\usepackage[ruled,vlined]{algorithm2e}
\usepackage{enumitem} 

\usepackage{tikz}
\usetikzlibrary{shapes,arrows}
\usetikzlibrary{intersections}
\usepackage{pgfplots}
\usepackage{pgfplotstable}
\usepackage{caption}
\usepackage{subcaption}
\usepackage{numprint}
\usepackage{textpos}
\usepackage{framed}

\begin{document}
\title{Trustworthy Pre-Processing of Sensor Data in Data On-chaining Workflows for Blockchain-based IoT Applications}
\titlerunning{Trustworthy Pre-Processing of Sensor Data in Data On-chaining Workflows }
\author{Jonathan Heiss \and Anselm Busse \and Stefan Tai}
\authorrunning{J. Heiss, A. Busse, and S. Tai }
\institute{Information Systems Engineering (ISE)\\
TU Berlin, Germany \\
\email{\{jh,ab,st\}@ise.tu-berlin.de}}
\maketitle              %

\begin{abstract}
    Prior to provisioning sensor data to smart contracts, a pre-processing of the data on intermediate off-chain nodes is often necessary. When doing so, originally constructed cryptographic signatures cannot be verified on-chain anymore. This exposes an opportunity for undetected manipulation and presents a problem for applications in the Internet of Things where trustworthy sensor data is required on-chain. 
    
    In this paper, we propose trustworthy pre-processing as enabler for end-to-end sensor data integrity in data on-chaining workflows. We define requirements for trustworthy pre-processing, present a model and common workflow for data on-chaining, select off-chain computation utilizing  Zero-knowledge Proofs (ZKPs) and Trusted Execution Environments (TEEs) as promising solution approaches, and discuss both our proof-of-concept implementations and initial experimental, comparative evaluation results. The importance of trustworthy pre-processing and principle solution approaches are presented, addressing the major problem of end-to-end sensor data integrity in blockchain-based IoT applications.
\end{abstract}
\section{Introduction}
\label{sec:introduction}
\input{contents/01_introduction.tex}
\section{Pre-Processing}
\label{sec:trustworthyProcessing}
\input{contents/03_TrustworthyPreprocessing.tex}
\section{Application}
\label{sec:concepts}
\input{contents/04_Application.tex}

\section{Evaluation}
\label{sec:evaluation}
\input{contents/05_Evaluation.tex}
\section{Discussion}
\label{sec:discussion}
\input{contents/06_Discussion.tex}

\section{Related Work}
\label{sec:relatedWork}
\input{contents/07_RelatedWork.tex}

\section{Conclusion}
\label{sec:conclusion}

\input{contents/08_Conclusion.tex}

\bibliographystyle{splncs04}
\bibliography{references}
\end{document}

%% file: contents/01_introduction.tex
Blockchain technology is increasingly used in the Internet of Things (IoT) to store and process critical sensor data originating from and shared between multiple, often mutually distrusting parties~\cite{ethertwin_Putz_2021,SupplyChain_Ikea_2020,realtimeIoT_finland_2020,DigitalTwin_beiking_2020,nettingPaper_eberhardt_2020,Healthcare_PatientMonitoring_2018,provenance_sigwart_2019}.  
In local energy grids with blockchain-based energy trading, for example, energy consumers and producers depend on smart meter-generated measurement data~\cite{bloGPV_useCase_2021,nettingPaper_eberhardt_2020}. In supply chains, product-related manufacturing and shipping events are written to a blockchain to provide a single source of truth for all involved, independent parties~\cite{SupplyChain_Ikea_2020,provenance_sigwart_2019}. In healthcare, blockchain use cases exist for doctors, hospitals, and emergency services to have access to patients' health data collected by wearables~\cite{Healthcare_PatientMonitoring_2018}. 

However, the variety and scale of connected IoT devices and the generated data pose new challenges regarding data processing and data on-chaining. Raw sensor measurements cannot directly be used on the blockchain because of volume limitations~\cite{IoTblockchain_storageCost_2020} or because sensitive information may be exposed and become accessible to unintended readers~\cite{nettingPaper_eberhardt_2020}. Blockchains inherently have privacy and scalability limitations~\cite{off-chaining_models_heiss,paper_eberhardt_tai_offchaining_patterns} that must be taken into account.

Consequently, the on-chain processing of sensor data is preceded by pre-processing steps to reduce data volume and ensure that confidential information is veiled. Such pre-processing typically is executed on intermediate, off-chain nodes as part of multi-staged data provisioning workflows~\cite{ethertwin_Putz_2021,SupplyChain_Ikea_2020,realtimeIoT_finland_2020,DigitalTwin_beiking_2020,nettingPaper_eberhardt_2020,Healthcare_PatientMonitoring_2018}: data originates on constrained sensor nodes, then moves to more powerful \emph{gateway} nodes for pre-processing, and is finally provisioned to smart contracts as  aggregated information.   
For example, in the healthcare use case described in~\cite{Healthcare_PatientMonitoring_2018}, data is pre-processed by personal computers or smartphones; in energy grids~\cite{nettingPaper_eberhardt_2020} by workstations located within participating households; in supply chains~\cite{SupplyChain_Ikea_2020} by board computers and mobile devices. 

While pre-processing has become an integral element in such \emph{data on-chaining workflows} and is necessary to mitigate scalability and privacy issues, off-chain pre-processing also represents a security risk. Sensor devices typically sign their measurements to provide data integrity. However, sensor data integrity is not end-to-end: once data is pre-processed on middleboxes, signatures constructed on the input do not apply to the output anymore. Contrary to smart contract application logic, application stakeholders cannot validate off-chain processing as part of the blockchain's consensus protocol.
Consequently, naive pre-processing can be exploited for malicious data manipulation without being noticed. This attack vector threatens data integrity in data on-chaining workflows and quickly questions the entire blockchain-based IoT system design and data quality.

To address this problem, solutions are needed to ensure \emph{trustworthy pre-processing}, i.e., to make computational correctness verifiable on the blockchain. 
Off-chain computations have been proposed~\cite{off-chaining_models_heiss} to outsource blockchain transaction processing to off-chain nodes without compromising trust guarantees.
Zero-Knowledge (ZK) computations and Trusted Execution Environments (TEE) are two important approaches here that are also increasingly being used in early-adoption projects and practice~\cite{nettingPaper_eberhardt_2020,zokVehicles_florida_2019,mediatingTrustworthiness_TEE_IoT_2020,DecIoTDataMngmnt_TEE__texas_2018}.
However, using ZK computations and TEEs for trustworthy pre-processing has not been examined so far.

In the face of the rising interest in blockchain-based sensor data management and the need for end-to-end sensor data integrity, in this paper, we analyze the underlying problem of trustworthy pre-processing in data on-chaining workflows, propose a model for integrity-preserving data on-chaining, and examine its practical applicability based on ZK computations and TEEs. 
Thereby, we make two individual contributions:

\begin{enumerate}
    \item First, we propose a model for end-to-end sensor data integrity through trustworthy pre-processing. We characterize sensor data pre-processing in on-chaining workflows for blockchain-based IoT applications based on relevant literature. From our findings, we refine our problem statement and introduce trustworthy pre-processing as a workflow element that enables application stakeholders through participation in the blockchain network to verify data integrity from source to sink. 
    \item Second, we examine the applicability of zkSNARKs-based and Trusted Execution Environments (TEE)-based off-chain computations for our proposed model. Based on a typical application workflow, we first conceptualize how trustworthy pre-processing can be instantiated with ZoKrates~\cite{paper_eberhardt_zokrates}, a toolkit for zkSNARKs-based off-chain computation, and with Intel SGX~\cite{IntelSGXExplained_Costan_2016}, Intel's realization of TEEs. Then, we implement the proposed model with both technologies as a proof of concept and present preliminary experiments in a testbed. While our results attest to the applicability of trustworthy pre-preprocessing with both approaches, they also confirm that, in comparison, zkSNARKs provide stronger integrity guarantees (weaker trust assumptions), whereas TEEs enable more efficient off-chain pre-processing.
\end{enumerate}

%% file: contents/03_TrustworthyPreprocessing.tex
To lay the foundation for trustworthy pre-processing, in this section, we first describe the general characteristics of pre-processing in blockchain-based IoT applications that we observed in pertinent research papers. Next, we refine our problem statement and define computational integrity, based on  ~\cite{ComputationalIntegrity_BenSasson_2017}.
Finally, we present a model for trustworthy pre-processing on gateway nodes for use in data on-chaining workflows that start with sensor devices and result in smart contracts. 

\subsection{Characterization}
\label{subsec:preprocessingCharacteristics}
Pre-processing in blockchain-based applications shares common objectives, input types, and functionality. 

\subsubsection{Objectives}
In data on-chaining workflows, off-chain pre-processing helps to mitigate blockchain-inherent scalability and privacy limitations. Thereby, it pursues the following objectives:  

\begin{itemize}
    \item \emph{Offloading Computation}: Outsource on-chain data processing to an off-chain node that is not bound to costly consensus-based transaction processing~\cite{nettingPaper_eberhardt_2020}.
    \item \emph{Reducing Storage}: Reduce the volume of sensor data to minimize the storage footprint on the blockchain~\cite{Healthcare_PatientMonitoring_2018,IoTblockchain_storageCost_2020}.
    \item \emph{Enabling Confidentiality}: Hide sensitive information contained in raw measurements or meta-data from stakeholders that do have read permissions~\cite{nettingPaper_eberhardt_2020,SupplyChain_Ikea_2020,ethertwin_Putz_2021}.
\end{itemize}

\subsubsection{Inputs}
Pre-processing can be executed on different types of data. We distinguish between the following: 
\begin{itemize}
    \item \emph{Measurements} include all data that is generated by sensor devices. This includes \emph{time series} data collected over a longer period of time~\cite{iotDataStorage_ETH_2017}, for example, temperature or location data, and \emph{event} data that represents externally triggered occurrences~\cite{SupplyChain_Ikea_2020}, for example, the scanning or opening of a container in a logistics context.
    \item \emph{Meta-data} originates from the sensor device and contains descriptive information about the measurements, such as sensor identities, target storage addresses, or timestamps. 
    \item \emph{Auxiliary data} is added at the gateway node. Examples are filter rules, access control lists, or storage addresses. 
\end{itemize}
Measurements and meta-data are critical for pre-processing and are referred to in the following as sensory data. 
In contrast, auxiliary data is never processed alone but optionally used to enrich pre-processing. 

\subsubsection{Types}
Without claiming completeness, we identify three general types of data pre-processing which can be observed in relevant applications~\cite{SupplyChain_Ikea_2020,bloGPV_useCase_2021,Healthcare_PatientMonitoring_2018,ethertwin_Putz_2021} and which represent typical functionality for operating on sequential data~\footnote{https://web.mit.edu/6.005/www/fa15/classes/25-map-filter-reduce/}.

\begin{itemize}
    \item \emph{Mapping}: Data is transformed into a target format, e.g., enumeration, encryption, decryption, hashing~\cite{ethertwin_Putz_2021,SupplyChain_Ikea_2020}.
    \item \emph{Reducing}: Data of one or multiple sensor devices is consolidated, e.g., the arithmetic average or a total amount is calculated~\cite{bloGPV_useCase_2021}.
    \item \emph{Filtering}: Data is filtered according to predefined rules, e.g., only values below a predefined threshold are returned~\cite{Healthcare_PatientMonitoring_2018}. 
\end{itemize}

\subsection{Problem Refinement}
\label{subsec:problemRefinement}
Data provisioning is often controlled by one of the stakeholders, e.g., shippers in supply chains~\cite{SupplyChain_Ikea_2020,realtimeIoT_finland_2020} or producers in energy markets~\cite{nettingPaper_eberhardt_2020}. 
Stakeholders may have a personal, often economically motivated interest in manipulating the data, e.g., in cooling chains to prevent contractual penalties if perishable fright is perished or to improve accounting positions.
Given such motifs, we assume data providing stakeholders as potential attackers. 

In data on-chaining workflows, data can take three states: it is \emph{in transit} when it is transmitted from one to another component, it is \emph{at rest} when it is persisted on disk, and it is \emph{in use} when it is processed in memory. 
During the states in transit and at rest, data integrity and authenticity can be verified using cryptographic signatures.
However, when data is processed, it is transformed and signatures constructed on the input do not apply for the output anymore. 
Furthermore, off-chain pre-processing cannot be validated by stakeholders through the consensus mechanism. 
An attacker could selfishly execute different functions on the data to manipulate the output and obtain a personal benefit without being noticed. 
Therefore, we assume manipulation of computation as the potential attack.

\subsection{Computational Integrity}
\label{subsec:compuational_integrity}
As a first step towards trustworthy pre-processing, we characterize computational integrity. We adopt the model proposed in~\cite{ComputationalIntegrity_BenSasson_2017}. 

A pre-processing program \emph{P} is executed on input data \emph{D} and some auxiliary data \emph{A} and returns output \emph{O} such that $P(D, A) \to O$. 

A malicious executer may benefit from creating a manipulated program $P'$ such that $P'(D, A) \to O' \: | \: O' \neq O$.
For example, in the supply chain use case, a shipper executes a threshold check $P$ on temperature measurements $D$ using the threshold $A$. If the shipper knows that the outcome $O$ triggers a contractual penalty, but $O'$ does not, it may change $P$ to $P'$ to obtain $O'$ instead of $O$. 
It then reports $O'$ to the blockchain and is exempt from the penalty. 
Additionally, the executer may leave the program $P$ unchanged but manipulate the input data D such that $P(D', A) \to O' \: | \: D \neq D' \land O' \neq O$ or the auxiliary data A such that $P(D, A') \to O' \: | \:  A \neq A' \land O' \neq O$

To prevent both, program and input manipulation, stakeholders should be able to verify computational integrity which is only guaranteed if output $O$ is executed on the  right program $P$ and on the right input data $(D, A)$ such that $P(D, A) \to O \: | \: (P \neq P') \land (D \neq D') \land (A \neq A')$. 
Therefore, we assume that program $P$ also generates an \emph{evidence} $E$ that asserts computational integrity such that $P(D, A) \to (O,E)$. 
To enable third-party stakeholders to verify computational integrity, additionally, an asymmetric key pair is required: the evidence signed with the \emph{proving key} can be verified by any third party with the corresponding \emph{verification key}. The evidence and the evidence key pair represent the major artefacts for trustworthy pre-processing.

\subsection{End-to-End Data Integrity}
\label{subsec:end-to-end_data_integrity}
\begin{figure}[htbp] %
    \centering
    \includegraphics[width=1\columnwidth]{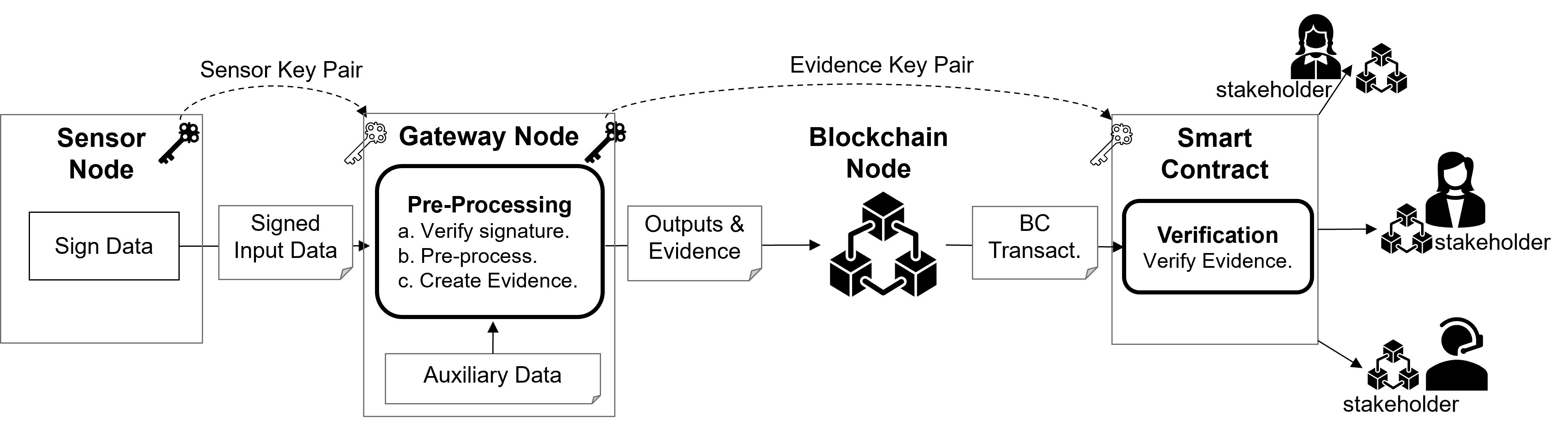}
    \caption{End-to-End Data Integrity through Trustworthy Pre-Processing}
    \label{fig:preprocessing}
\end{figure}

Given that integrity of data can be verified while it is in use, we can define a data on-chaining workflow where integrity is verifiable from its source on the sensor node to its sink on the smart contract as depicted in Figure~\ref{fig:preprocessing}. Note that instead of a simple signature, verifiable evidence is provided to the blockchain that allows data integrity verification with moderate computational overhead in the blockchain network.

\subsubsection{One Time Setup} 
During an initial one time setup, central system artifacts are generated and deployed on the system components. Given that these artifacts are critical to verify computational integrity, we assume a trusted setup where each stakeholder can verify the integrity of the artifacts.
It consists of three steps: 

As a first step (\emph{1. Integrity Assertion}), an environment is established that enables the gateway node to generate verifiable evidence of computational integrity as accompanying artefacts of the pre-processing outputs. This includes the integrity of sensory and auxiliary inputs. Examples for such environments are mathematical constraint systems~\cite{paper_eberhardt_zokrates} or trusted execution environments~\cite{IntelSGXExplained_Costan_2016} as will be described in the subsequent section. 

Next (\emph{2. Key Generation}), two key pairs are required: an \emph{evidence key pair} consisting of a proving and verification key for signing and verifying the evidence and a \emph{sensor key pair}, represented as a cryptographic public and private key that is used to sign and verify the sensor data on the sensor node and the gateway node respectively. 

As the last setup step (\emph{3. Deployment}), all artefacts are deployed: The gateway node is equipped with the sensor node's public key, the integrity-preserving pre-processing program, the proving key, and optionally auxiliary data. The smart contract receives the verification key that enables evidence verification. 

\subsubsection{Recurring Operations} 
Sensory data arrives recurringly at the gateway node in regular intervals, e.g., batches of \emph{time series} data, or in irregular intervals, e.g., externally triggered \emph{events}.
Then (\emph{4. Pre-Processing}), the pre-processing program takes the signed sensory data, the sensor's public key, and optionally auxiliary data as inputs and executes the following steps: 

\begin{enumerate}[label=(\alph*)]
    \item The sensory inputs' signature is verified with the sensor device's public key. 
    \item Pre-processing functions are executed on the verified inputs. Examples are provided in section~\ref{subsec:preprocessingCharacteristics}.
    \item An evidence is created and signed with the gateways' proving key. The evidence enables the smart contract to verify computational integrity. 
\end{enumerate}

Outputs and signed evidence are transmitted to the smart contract through the blockchain node.
The smart contract verifies the evidence using the verification key (\emph{5. Verification}). 
Successful verification on the blockchain enables applications stakeholders to independently verify that integrity of sensor data has been preserved from source to sink despite intermediate pre-preprocessing. 
Pre-processing outputs can be consumed through participating blockchain nodes and used for subsequent processing. 

\begin{figure}[htbp] %
    \centering
    \includegraphics[width=1\columnwidth]{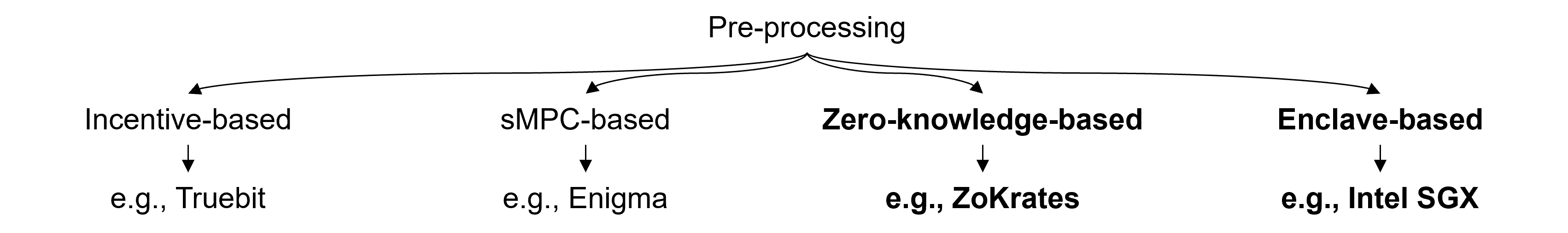}
    \caption{Off-chain Computation Technologies according to~\cite{off-chaining_models_heiss}}
    \label{fig:off-chain_computation}
\end{figure}

%% file: contents/04_Application.tex
For trustworthy pre-processing to become easily applicable in practice, technologies are required that enable on-chain verifiability of computational integrity and that can implement the pre-processing characteristics as described in~\ref{subsec:preprocessingCharacteristics}.

\subsection{Technologies for Trustworthy Pre-processing}
Off-chain computation has been proposed to mitigate privacy and scalability limitations of blockchains by outsourcing computation to off-chain nodes without compromising core blockchain properties~\cite{off-chaining_models_heiss,paper_eberhardt_tai_offchaining_patterns}. Thereby, it represents a matching concept for trustworthy pre-processing.

However, the different approaches to off-chain computation presented in~\cite{off-chaining_models_heiss} and depicted in Figure~\ref{fig:off-chain_computation} are not equally suitable. Both incentive-based and sMPC-based approaches require multiple nodes that execute non-trivial protocols. However, in data on-chaining applications in the IoT~\cite{Healthcare_PatientMonitoring_2018,realtimeIoT_finland_2020,SupplyChain_Ikea_2020,nettingPaper_eberhardt_2020,ethertwin_Putz_2021}, pre-processing is typically executed on a single node with limited networking and storage capacity. 
If such a constraint is given, the distributed computation model and interactive nature of incentive- and sMPC-based approaches may be inconsistent with use case specific requirements which restricts general applicability. %
In contrast, zero-knowledge and enclave-based approaches can be executed non-interactively on a single node and, hence, promise broader applicability for trustworthy pre-processing. 

\subsection{ZkSNARKs-based Pre-Processing with ZoKrates}
\label{subsec:zkSNARKsWorkflow}

\emph{Zero-knowledge proofs} enable a prover to convince a verifier that it has correctly executed a computation without revealing inputs to the verifier. 

\emph{zkSNARKs} can be summarized as one type of a zero-knowledge protocol that distinguishes through \emph{succinctness}, i.e., resulting artefacts are small in size and can be verified fast, \emph{non-interactivity}, i.e., only one message is required to convince the verifier, and \emph{argument of knowledge}, i.e., the prover is able to prove that she has access to the correct data.

ZoKrates~\cite{paper_eberhardt_zokrates} provides a toolbox and a higher-level language to implement a zkSNARKs-proving system where an off-chain prover can convince an on-chain verifier that the computation has been executed correctly. 

\begin{figure}[htbp] %
    \centering
    \includegraphics[width=1\columnwidth]{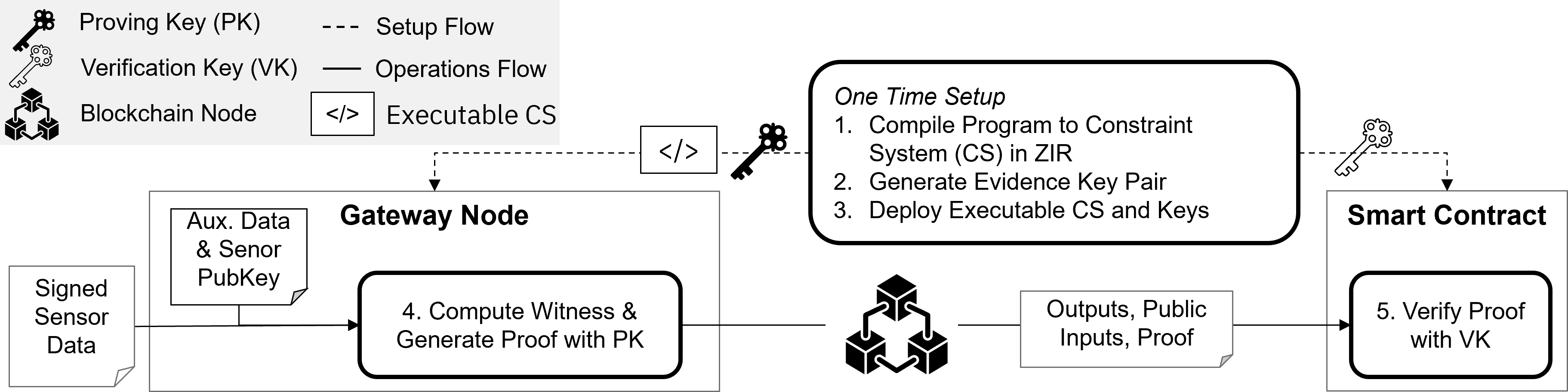}
    \caption{Trustworthy Pre-Processing with ZoKrates}
    \label{fig:zokrates_preprocessing}
\end{figure}

To describe the ZoKrates-based pre-processing (compare Figure~\ref{fig:zokrates_preprocessing}), we leverage the model presented in Section~\ref{subsec:preprocessingCharacteristics} and build upon the ZoKrates workflow described in~\cite{paper_eberhardt_zokrates}.

\subsubsection{One Time Setup}
\begin{enumerate}
    \item \emph{Integrity Assertion}: To guarantee integrity of auxiliary data and the sensor public key, both are typed as public arguments in the ZoKrates program and, hence, are required on-chain for evidence verification. Since the verification would fail on different public inputs, their integrity can be determined on-chain. 
    
    Once specified, the high-level ZoKrates code is compiled into an executable constraint system (ECS) in the ZoKrates Intermediate Representation (ZIR) format that can be considered as an extension to a Rank-1-Constraint System and enables assertion of computational integrity: if a variable assignment is found that satisfies the defined constraints computational integrity can be proven.
    \item \emph{Evidence Key Generation}: An evidence key pair is generated from a Common Reference String (CRS)~\cite{paper_eberhardt_zokrates} which enables proof creation and verification. Since the CRS allows construction of fake proofs it must be securely disposed after key generation. The evidence key pair is cryptographically bound to the previously generated ECS.
    \item \emph{Deployment}: The ECS, the evidence proving key, auxiliary data, and the sensor public key are deployed to the gateway node which takes the role of the off-chain prover. Verification key and the verification contract are deployed to the blockchain. %
\end{enumerate}

\subsubsection{Recurring Operations}
\begin{enumerate}
\setcounter{enumi}{4}
    \item \emph{Execution}: The ZIR program is executed on predefined inputs, through the ZoKrates interpreter. The output is called witness, an artefact representing variable assignments that satisfy the specified constraints for a specific execution. In a separate step, the cryptographic proof is generated based on the execution-specific witness and the program-specific proving key. Finally, outputs and evidence are forwarded to the smart contract through a blockchain node.
    \item \emph{Verification}: The verification contract takes the cryptographic proof, the verification key, and public program arguments as input parameters. The verification is only successful if the proof is executed with the right program and on the right (public) inputs.
\end{enumerate}

\begin{figure}[htbp] %
    \centering
    \includegraphics[width=1\columnwidth]{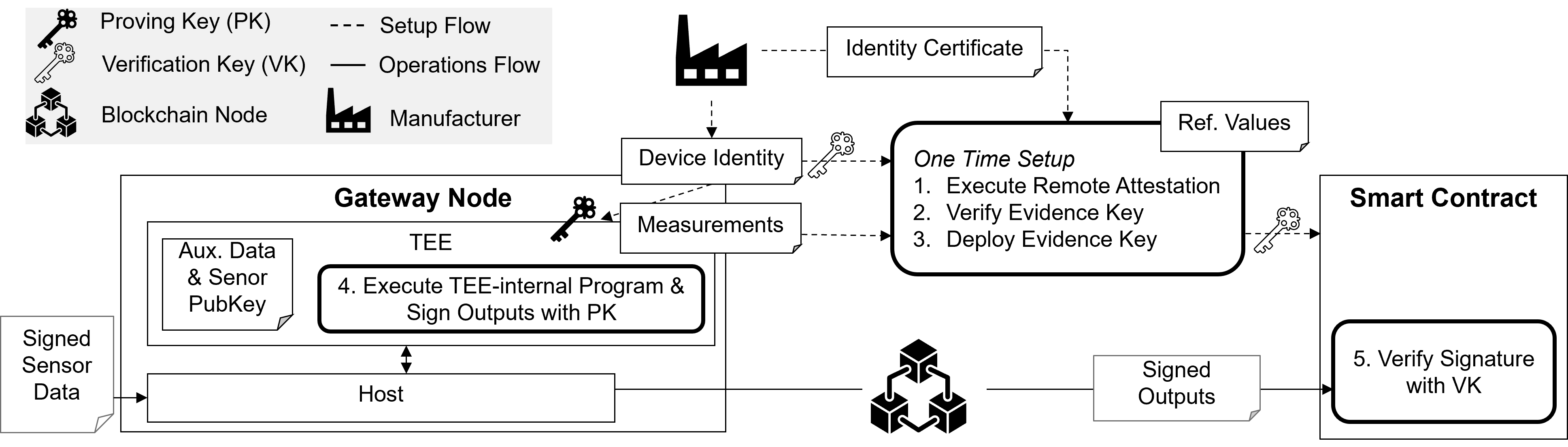}
    \caption{Trustworthy Pre-Processing with Intel SGX}
    \label{fig:enclave_preprocessing}
\end{figure}

\subsection{Enclave-based Pre-Processing with Intel SGX}
\label{subsec:enclaveWorkflow}
Enclave-based computation enables an enclave-external party to verify that an output has been computed by a specific program inside a specific enclave that protects internal integrity.
Thereby, it relies on two concepts: Trusted Execution Environments and Remote Attestation. 

\emph{Trusted Execution Environments} (TEE) are hardware-secured parts of a system architecture that protect data and code from external manipulation and disclosure. 
Programs executed inside such TEEs are running in an isolated and/or encrypted memory region that cannot even be accessed in the highest privilege level of the system.
Thus, it protects the content of the TEE from the system owner and guarantees the integrity of computation executed inside the TEE.
Intel SGX is Intel's concrete implementation of TEEs. We use the terms TEE and enclave interchangeably.

\emph{Remote Attestation} enables the external verification of the integrity of the TEE's internal state and the authenticity of messages received from inside.
Thus, ensuring that a malicious attacker cannot falsely pose as an trusted enclave.
TEE-enabled devices have a device identity key that is embedded into the device hardware during manufacturing and can be verified by external parties through a Public Key Infrastructure (PKI).
Using this key, the device creates for each instantiated TEE an identity certificate which can externally be verified through the PKI. This enables evidence key generation.  
When remote attestation is requested, the enclave returns signed measurements which represent a complete snapshot of the TEEs internal state.
With SGX as TEE, remote attestation and the PKI are managed by Intel.

In the following, we describe pre-processing with Intel SGX as depicted in Figure~\ref{fig:enclave_preprocessing}. To achieve comparability with Zokrates-based pre-processing we use the same workflow model as described in Section~\ref{subsec:end-to-end_data_integrity}.

\subsubsection{One Time Setup}
\begin{enumerate}
    \item \emph{Integrity Assertion}: To guarantee integrity of auxiliary data and the sensor public key, both must be protected through the TEEs security guarantees. Therefore, they are specified inside the enclave during implementation. 
    
    Once the enclave is instantiated and loaded in memory, as a first step, remote attestation is executed to verify the enclave's internal state. The signed measurements are verified using the enclave's public key that is previously authenticated through the externally managed PKI. If the measurements match a predefined reference value that represents the ground truth of the enclave's internal state, the enclave's integrity is verified. 
    \item \emph{Key Generation}: To verify the enclave's integrity a unique enclave-bound key pair is required that can be authenticated from outside the enclave. This evidence key pair is used to sign program results computed inside the enclave. Given that the enclave's integrity guarantees hold, this signature enables verification of computational integrity on the blockchain. The evidence key is generated inside the enclave and can be authenticated through an externally managed PKI. 
    \item \emph{Deployment}: The enclave's evidence public key becomes part of the verification contract which implements the signature verification on-chain and is deployed to the blockchain. At this point, the enclave is already instantiated on the gateway node. 
\end{enumerate}

\subsubsection{Recurring Operations}
\begin{enumerate}
    \setcounter{enumi}{4}
    \item \emph{Execution}: Sensor data is provided through the host program which represents the only interface to the enclave. Auxiliary data and the sensor public key are already part of the enclave and, hence, protected. 
    The program is executed as defined in Section~\ref{subsec:end-to-end_data_integrity}. The computational outputs are signed with the evidence proving key.
    \item \emph{Verification}: The verification contract validates the signature with the evidence verification key. A successful validation proves the outputs' authenticity, i.e., they have been signed with the right proving key that is unique to the enclave, and integrity, i.e., the received outputs are computed by the right pre-processing program inside the enclave.
\end{enumerate}

%% file: contents/05_Evaluation.tex
Given the two conceptual workflow descriptions, in this section, we evaluate the technical feasibility for each technology. 

\subsection{Implementation}
\label{subsec:implementation}
Our proof-of-concept (PoC) implementations follow the descriptions provided in Section~\ref{subsec:zkSNARKsWorkflow} and~\ref{subsec:enclaveWorkflow} respectively. %
Thereby, we focus on the recurring operations steps, \emph{execution} and \emph{verification} which we consider as most relevant to demonstrate feasibility.
Aspects of the setup phase are discussed in Section~\ref{sec:discussion}.

The PoC program should respect the pre-processing characteristics presented in Section~\ref{subsec:preprocessingCharacteristics}. 
Our program mimics a threshold violation check on sensory data where the threshold represents auxiliary data. 
The sensory data is \emph{filtered} for violations, then \emph{reduced} by counting the violations, and \emph{mapped} by scaling the filtered values down. The smart contract is only provided with the violation count. 
Thereby, the program fulfills all three objectives: computation is outsourced to an off-chain node, the data footprint is reduced in size, and the potentially sensitive sensor measurements are not published on-chain.

\textbf{ZoKrates:}
For our ZoKrates-based implementation, we simulate the sensor node with a Python script that hashes the data with SHA256 and signs it with EDDSA-based sensor key pair, which ZoKrates support. 
Plain sensory data is a private input, while the data's hash, signature, and the sensor public key are public inputs to the ZoKrates program. 
To verify integrity of sensory inputs, the signature's hash input is reconstructed from the plain sensor data and compared to the hash inputs. Only if both signature verification and hash comparison are successful integrity is guaranteed. 
Hashing and signature verification are implemented using the ZoKrates Standard Library. 
Pre-processing is executed by two commands provided by the ZoKrates CLI: \emph{compute-witness} that requires the compiled program and \emph{generate-proof} that takes proving key and witness as inputs. The outputs are written to disk.

\textbf{Intel SGX:}
For the SGX evaluation, we have implemented two enclaves. 
The first one simulates a sensor node and signs the sensory input data with an internally generated sensor key pair using the SGX-provided operations \emph{sgx\_create \_keypair} and \emph{sgx\_ecdsa\_sign}.
The second enclave represents the gateway node that stores auxiliary data and the sensor public key internally.
It verifies the sensor data with the sensor public key using the SGX operation \emph{sgx\_ecdsa\_verify}. 
Evidence key pair generation and signature construction on computational outputs are realized with the same SGX commands as the sensor enclave.
The processing result and the corresponding signature are written to disk.

\textbf{Ethereum:} As blockchain technology, we chose Ethereum~\cite{paper_wood_ethereum_yellow}, which is widely used and finds application both as a public blockchain but also as consortium blockchain based on Proof-of-Authority consensus and non-public deployment. 
For each, respectively, a verification contract is implemented in Solidity that runs on a locally deployed Ethereum blockchain and is accessed through a Ganache blockchain client. 
To validate Intel SGX evidence, we build upon an existing ECDSA implementation for the Ethereum blockchain~\footnote{https://github.com/tdrerup/elliptic-curve-solidity}.
ZoKrates proofs rely on EdDSA (twisted Edwards curve) and are verified through a dedicated verification contract that is generated by ZoKrates CLI support~\footnote{https://github.com/Zokrates/ZoKrates}.

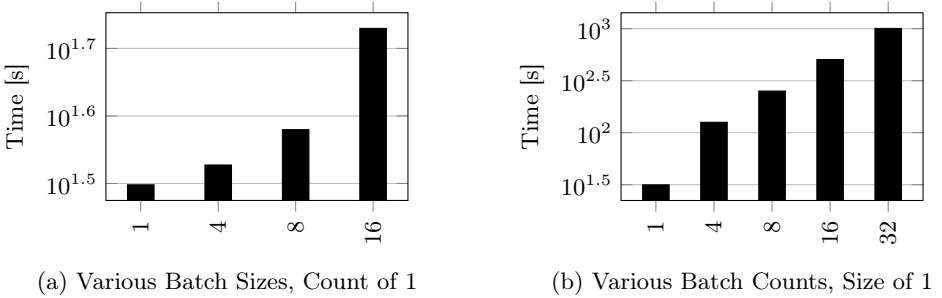
\begin{figure}[htbp]
	\begin{minipage}[t]{0.5\textwidth}
	\pgfplotstableread{
1 31.49
4 33.69
8 38.01
16 53.68
32 0.0
}\nsdata%
\begin{tikzpicture}
\begin{axis}[
    ybar=1pt,
    bar width=10pt,
    xtick={0,1,2,3,4,5,6,7,8,9,10},
    xticklabels from table={\nsdata}{[index]0},
    x tick label style={
        /pgf/number format/1000 sep=
    },
    enlarge x limits=0.15,
    ylabel={Time [s]},
    ylabel near ticks,
    legend style ={at={(0.02,0.97)}, anchor=north west, legend columns=2},
    legend cell align={left},
    xmin=0,
    width=4cm,
    height=2.5cm,
    scale only axis,
    xmode=normal,
        ymode=log,
       log basis y={10},
                xticklabel style={rotate=90},
       ymajorgrids,
      ymin=0
]
\addplot[fill=black] table[x expr=\coordindex,y index=1] {\nsdata};
\end{axis}
\end{tikzpicture}
\subcaption[]{Various Batch Sizes, Count of 1}
\end{minipage}
\qquad%
	\begin{minipage}[t]{0.5\textwidth}
\pgfplotstableread{
1 31.487
4 125.892
8 251.398
16 505.078
32 1006.261
}\nsdata
        \begin{tikzpicture}
\begin{axis}[
    ybar=1pt,
    bar width=10pt,
    xtick={0,1,2,3,4,5,6,7,8,9,10},
    xticklabels from table={\nsdata}{[index]0},
    x tick label style={
        /pgf/number format/1000 sep=
    },
    enlarge x limits=0.15,
    ylabel={Time [s]},
    ylabel near ticks,
    legend style ={at={(0.02,0.97)}, anchor=north west, legend columns=2},
    legend cell align={left},
    xmin=0,
    width=4cm,
    height=2.5cm,
    scale only axis,
    ytick={31.6,100,316,1000},
    xmode=normal,
        ymode=log,
       log basis y={10},
                xticklabel style={rotate=90},
       ymajorgrids
]
\addplot[fill=black]  table[x expr=\coordindex,y index=1] {\nsdata};
\end{axis}
\end{tikzpicture}%
\subcaption[]{Various Batch Counts, Size of 1}
\end{minipage}
    \caption{Pre-processing with ZoKrates}
    \label{fig:resultsZokrates}
\end{figure}

\subsection{Experiments}
\label{subsec:experiments}
Given our proof-of-concept implementations, we can now conduct initial experiments to obtain the first practical insights into trustworthy pre-processing with zkSNARKs and TEEs.
At this point, it should be noted that experimental results strongly depend on our non-optimized PoC implementations and, hence, cannot simply be generalized. 

\subsubsection{Exerimental Setup}
For our experimental setup, we deploy our implementations on an Intel NUC-Kit NUC7PJYH with an SGX enabled Pentium Silver J5005 CPU, 8 GB of Memory, and an Ubuntu 18.04.5 LTS operating system. 
To construct workloads, we use smart meter measurements collected in a testbed of an energy grid research project\footnote{https://blogpv.net/} and prepare the measurements such that (1) each measurement consists of four integer values, (2) measurements are collected into batches of different sizes line-wise in plain text, and (3) each batch is signed to represent the sensor's signature.

As mentioned in Section~\ref{subsec:preprocessingCharacteristics}, pre-processing is typically exposed to two types of workloads: event and batch processing. To simulate that in our experimental setup, we turn on two knobs: for events of different sizes, we change the input data size per execution (batch size), for batch processing, we vary the number of subsequent executions (batch count). Latter is executed on size-one-batches which contain a single measurement.

The computational outputs of size-one-batch experiments are used for on-chain verification, which is measured in \emph{Gas}, an Ethereum-specific metric for capturing computational complexity of on-chain transaction processing.

\subsubsection{Results}
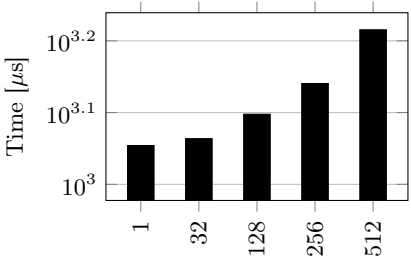
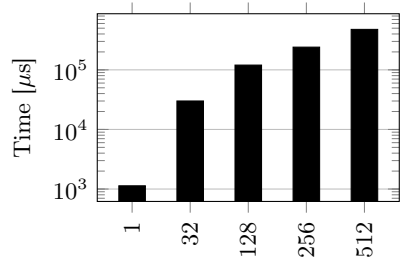
\begin{figure}[htbp]
	\begin{minipage}[t]{0.5\textwidth}
\pgfplotstableread{
1 1133.15
32 1158.4
128 1252.97
256 1382.66
512 1642.7
}\nsdata%
\begin{tikzpicture}
\begin{axis}[
    ybar=1pt,
    bar width=10pt,
    xtick={0,1,2,3,4,5,6,7,8,9,10},
    xticklabels from table={\nsdata}{[index]0},
    x tick label style={
        /pgf/number format/1000 sep=
    },
    enlarge x limits=0.15,
    ylabel={Time [$\mu$s]},
    ylabel near ticks,
    legend style ={at={(0.02,0.97)}, anchor=north west, legend columns=2},
    legend cell align={left},
    xmin=0,
    width=4cm,
    height=2.5cm,
    scale only axis,
    xmode=normal,
        ymode=log,
       log basis y={10},
                xticklabel style={rotate=90},
       ymajorgrids,
      ymin=950
]
\addplot[fill=black]  table[x expr=\coordindex,y index=1] {\nsdata};
\end{axis}
\end{tikzpicture}
\subcaption[]{Various Batch Sizes, Count of 1}
\end{minipage}
\qquad%
\begin{minipage}[t]{0.5\textwidth}
\pgfplotstableread{
1 1133.15 1138.22 1382.66 1642.7
32 30034.28 30535.35 37901.16 45501.82
128 120232.71 121998.32 149911.69 180068.11
256 240551.12 243644.98 300823.25 359017.89
512 477618.48 487484.42 599770.0 721393.99
}\nsdata

        \begin{tikzpicture}
\begin{axis}[
    ybar=1pt,
    bar width=10pt,
    xtick={0,1,2,3,4,5,6,7,8,9,10},
    xticklabels from table={\nsdata}{[index]0},
    x tick label style={
        /pgf/number format/1000 sep=
    },
    enlarge x limits=0.15,
    ylabel={Time [$\mu$s]},
    ylabel near ticks,
    legend style ={at={(0.02,0.97)}, anchor=north west, legend columns=2},
    legend cell align={left},
    xmin=0,
    width=4cm,
    height=2.5cm,
    scale only axis,
    xmode=normal,
        ymode=log,
       log basis y={10},
                xticklabel style={rotate=90},
       ymajorgrids
]
\addplot[fill=black]  table[x expr=\coordindex,y index=1] {\nsdata};
\end{axis}
\end{tikzpicture}%
\subcaption[]{Various Batch Counts, Size of 1}
\end{minipage}
    \caption{Pre-processing with Intel SGX}
    \label{fig:resultsSGX}
\end{figure}

The results summarized for ZoKrates in Figure~\ref{fig:resultsZokrates} and for Intel SGX in Figure~\ref{fig:resultsSGX} show the overall execution time for off-chain pre-processing in seconds and microseconds, respectively. 
As expected, the execution time of zkSNARKs-based pre-processing is orders of magnitude higher than that of enclave-based pre-processing. 
With larger batch sizes, the execution time increases almost gradually. This holds true for each technology individually as shown in Figure~\ref{fig:resultsZokrates} a) and Figure~\ref{fig:resultsSGX} a).
Similar behaviour can be observed for increasing the batch count as shown in Figure ~\ref{fig:resultsZokrates} b) and Figure~\ref{fig:resultsSGX} b). However, we can observe that for both ZoKrates and SGX the increase is much steeper for a growing batch count than for a growing batch size (note the different logarithmic y-scales). For this specific implementation example, this would mean that it is preferable to increase the number of processed data through larger batch sizes rather than counts when possible in the actual application scenario.

In ZoKrates-based pre-processing, the accompanying construction of cryptographic proofs represents a memory-intensive computation that correlates with the input size. 
The experiment for the next larger batch size of 32~measurements in ZoKrates ran out of memory during the \emph{proof-generation} on the test system. 
Given that sensory data can quickly grow very large, the memory capacity of constrained IoT or edge devices may present a limiting factor, but may not be an issue for larger middleboxes.

In contrast, Intel SGX reduces pre-processing overhead. Even though, our implementation was also memory limited regarding a batch size larger than 1024~measurements, this is just a limitation of the current SGX design that might change in the future and can be mitigated, e.g., by splitting up the processes into multiple enclaves on the same machine. Better efficiency and smaller memory consumption distinguishes Intel SGX as a suitable technology for lower IoT layers where computational resources are typically scarce. However, contrary to ZoKrates, SGX-based pre-processing requires an increased trust in the correctness of the hardware implementation and the attestation process that requires trusting Intel regarding a correct attestation.

In our proof-of-concept implementation, on-chain verification costs are cheaper for ZoKrates-generated proofs (\numprint[Gas]{567614}) than for Intel SGX-generated signatures (\numprint[Gas]{1211443}). However, since on-chain verification costs strongly depend on the implementation of respective signature algorithm our results cannot be generalized, e.g., for other blockchain technologies.

%% file: contents/06_Discussion.tex
While in the previous section, initial insights about the performance behavior of each technology were provided, in this section, we discuss security and trust aspects and potential extensions for trustworthy pre-processing.

\textbf{Integrity and Trust Assumptions:} 
As described in Section~\ref{subsec:problemRefinement}, pre-processing is assumed to be executed by non-trusted stakeholders who have an incentive for data manipulation. While off-chain technologies eliminate unnoticed attacks during pre-processing, the setup phase still reveals an attack surface. 
In Zokrates, for example, key generation must be executed in a trusted setup to guarantee that the Common Reference String is safely disposed to prevent fake proof generation. However, establishing a trusted setup for zkSNARKs is a known problem to which various approaches exist as referenced in~\cite{paper_eberhardt_zokrates}.
In Intel SGX, the integrity guarantee strongly relies on the internal state of the enclave and on the authenticity of the evidence key pair. To preserve this guarantee, remote attestation and key authenticity must be verified through a trusted third party or by all involved stakeholders individually. Also, auxiliary data and the sensor's public key must be verified before being added to the enclave.
Beyond the setup, zkSNARKs-based pre-processing does not rely on further trust assumptions, whereas enclave-based pre-processing heavily relies on a trustworthy manufacturer that ensures that private keys are kept secret and certificates obtained from the PKI are authentic to the device's identities. This distinguishes ZoKrates as particularly suitable for processing critical data with substantial security demands. 

\textbf{Further Attacks:}
Beyond our attack model described in Section~\ref{subsec:problemRefinement}, attacks on data freshness and availability must be considered. 
While an attacker that controls communication channels, e.g., between gateway and blockchain node, cannot compromise data integrity without being noticed (\emph{Man-in-the-Middle Attack}) due to signature and evidence verification, it can, however, intercept and replay messages in a different order to impact the overall application logic (\emph{Replay Attack}). To prevent this, secure timestamps or challenge-response patterns can be applied.
Furthermore, to prevent a malicious executor from compromising availability by withholding messages (\emph{Denial of Service Attack}), gateway nodes can redundantly be deployed to eliminate centralization, similar to this proposal~\cite{dior_distributedTEEOracle_korea}.

\textbf{Multi-Stage Pre-Processing:} 
In multi-stage data on-chaining workflows, multiple pre-processing tasks may be executed subsequently by different non-trusted stakeholders. To verify integrity on-chain, an evidence chain must be established that allows any subsequent computation to validate the provided evidence of the previous computation. This way, end-to-end integrity could be guaranteed along arbitrarily long on-chaining workflows. 

\textbf{Confidential Pre-Processing:} 
While this work focuses on integrity preservation, in some use cases it might be required to keep inputs to pre-processing hidden from the executor. This can, for example, be achieved through Intel SGX, where encrypted inputs can be decrypted inside the enclave, processed, and encrypted again before being returned. Thereby, inputs and outputs would not be accessible by the executor. However, side-channel attacks must be respected that are known to extract confidential information from enclaves~\cite{foreshadow:2018}.

%% file: contents/07_RelatedWork.tex
In this paper, we extend trustworthy data on-chaining as presented in~\cite{trustworthyOnchaining_heiss} by considering \emph{data in use} as an additional attack vector. Furthermore, we leverage approaches to off-chain computation presented in~\cite{off-chaining_models_heiss} to realize trustworthy pre-processing. 
From the proposed off-chain computation technologies in~\cite{off-chaining_models_heiss}, zkSNARKs and Trusted Execution Environments are increasingly adopted in scientific literature on blockchain-based IoT applications. 

Recently, many proposals leverage zkSNARKs for off-chain computations through Zokrates; however, only a few intersect blockchain-based sensor data management. 
While in~\cite{nettingPaper_eberhardt_2020} ZoKrates is applied for off-chain processing of sensor data, i.e., smart meter measurements in local energy grids, other works mainly use Zokrates for privacy-preserving authentication, e.g. in the context of smart vehicle authentication at charging stations~\cite{zokVehicles2_florida_2020}, consumer authentication for car sharing~\cite{ZokCarSharing_TUDresden_2020}, or in health care for patient authentication~\cite{zokHealthcare_indian_2020}. 

TEEs are leveraged in various papers to implement trustworthy oracles that bridge data provisioning from off-chain data sources to smart contracts. For example, in TownCrier~\cite{TownCrier}, a TEE-based oracle system is proposed to authenticate data provided by HTTPS-enabled off-chain data sources, or in~\cite{dior_distributedTEEOracle_korea}, a distributed TEE-enabled oracle system is proposed that improves availability.
Beyond scientific usage, e.g., ChainLink~\footnote{https://chain.link/} works on a solution to implement these concepts for practical usage~\cite{chainlink2_breidenbach_2021}.

While the main focus of these proposals lies in data provisioning, other works instead use TEEs for sensor data management. 
In~\cite{mediatingTrustworthiness_TEE_IoT_2020}, for example, a system is proposed that employs TEEs for intermediate processing of sensory data before it is forwarded to the blockchain and the cloud. 
The authors of~\cite{DecIoTDataMngmnt_TEE__texas_2018} use TEEs for trustworthy access management of sensor data in hybrid storage systems where off-chain storage holds encrypted sensor data and the blockchain stores its hashes and access logs.
While these proposals do not apply pre-processing as defined in this paper, they underline the need for a systematization of trustworthy pre-processing that we aim to provide with our contributions.

%% file: contents/08_Conclusion.tex
End-to-end sensor data integrity is critical to many blockchain-based IoT applications. Data on-chaining workflows accordingly require pre-processing on off-chain nodes to be trustworthy. 
In this paper, we explored the use of zkSNARKs- and TEE-based computations for trustworthy pre-processing, first, as individual candidate technologies that require non-trivial set-ups for integration in data on-chaining workflows, and second, through a preliminary, comparative experimental evaluation based on two proof-of-concept implementations. 
We conclude that each presents an important approach that (a) can conceptually be well-integrated in respective workflows and (b) satisfies the requirements and primary objective of end-to-end data integrity. 
Our proof-of-concept implementations use current, state-of-the-art software, and, since both zero-knowledge proofs and TEEs are very active areas of research, our implementations and the experimental findings must be seen as preliminary. 
We expect rapid advances regarding the used software stacks and current constraints regarding memory limitations, and, consequently, performance numbers to change. Still, a principal performance gap and performance advantage of TEEs over zkSNARKs is expected to remain. 
However, as discussed in this paper, the choice of an approach and technology will depend also on other, non-performance criteria like the integrity and trust assumptions or existing attack vectors for the specific IoT application under consideration.
Future work will address extensions of the proposed model regarding its computational scalability through parallel execution and its applicability for stream processing.